\documentclass[prl,superscriptaddress,twocolumn,showpacs]{revtex4}
\usepackage{amssymb,amsmath}
\usepackage{color}
\usepackage{dsfont}
\usepackage{hyperref}
\usepackage{graphicx}

\definecolor{darkblue}{rgb}{0,0,0.5}
\definecolor{darkred}{rgb}{0.5,0,0}

\hypersetup{
  pdfborder={0,0,0},
  colorlinks=true,
  linkcolor=darkblue,
  citecolor=blue
}

\begin{document}

\title{Continuous-Variable Bell Inequalities in Phase Space}

\author{Karl-Peter Marzlin}
\affiliation{Department of Physics, St. Francis Xavier University,
  Antigonish, Nova Scotia, B2G 2W5, Canada}
\affiliation{Institute for Quantum Information Science,
        University of Calgary, Calgary, Alberta T2N 1N4, Canada}

\author{T.~A.~Osborn}
\affiliation{Department of Physics and Astronomy, University of Manitoba,
Winnipeg, Manitoba R3T 2N2, Canada} \bigskip

\begin{abstract}
We propose a variation of Bell inequalities for continuous
variables that employs the Wigner function and Weyl symbols of
operators in phase space.
We present examples of Bell inequality
violation which beat Cirel'son's bound.
\end{abstract}


\maketitle

Since its discovery in 1964 \cite{Bell:Physics1964} the Bell
inequality (BI)
has triggered an enormous interest in the differences
between classical and quantum correlations.
Bell inequalities are now commonly referred to as equations
that relate correlation measurements which
are fulfilled by any local hidden variable (LHV) theory,
but are violated within the framework of quantum mechanics (QM).
The original inequality
was formulated for dichotomic variables in spin systems.
Clauser, Horne, Shimony, and Holt (CHSH)
\cite{PhysRevLett.23.880} presented a BI that
was more amenable for experimental tests and is
nowadays widely used.

The original BI was inspired by
the Einstein-Podolsky-Rosen paradox
\cite{PhysRev.47.777,RevModPhys.81.1727} for continuous variables (CV).
BIs for CV
systems were first developed using dichotomic variables that have
eigenvalues $\pm 1$ \cite{JModOpt42-939,Gour2004415,Praxmeyer:EPJD2005} . Recently,
a new approach for CV Bell inequalities has been
developed by Cavalcanti, Foster, Reid, and Drummond (CFRD)
\cite{PhysRevLett.99.210405,arXiv:1005.2208}. The CFRD inequality
can be formulated for arbitrary CV observables, but a multi-partite
quantum system with five or more spatially separated sub-systems
is usually required to obtain a violation.
In this paper we offer a generalization of CFRD inequalities that
employs the Wigner function $ W(q,p)$.
The resulting BIs are conceptually
different from those obtained using operator
methods. Furthermore, the
similarity  of quantum phase space expectation values to those of
classical statistical mechanics can provide new insight to the BI.

The Wigner function is a quasi-probability distribution
in phase space that is defined (in a 1D system) by \cite{PhysRev.40.749}
$ W(q,p) = (2\pi \hbar)^{-1}  \text{Smb}[\hat{\rho}](q,p)$, with
\begin{align}
  \text{Smb}[\widehat{A}](q,p) &=2
  \int_{-\infty}^\infty  \; \langle q-q'| \widehat{A} |  q+q' \rangle
   e^{2i p q'/\hbar} \,dq'
\label{eq:wignerfunction} \end{align}
the Weyl symbol of an operator $\widehat{A}$ and
$\hat{\rho}$ the density matrix. $ W(q,p)$
appears in the expectation values of $\widehat{A}$,
\begin{align}
  \langle \widehat{A} \rangle &=
  \int dq\, dp\, W(q,p)\, \text{Smb}[\widehat{A}](q,p)\; .
\label{eq:weylExpec}\end{align}
Phase space methods have been used to study specific
implementations of the CHSH inequality for dichotomic variables
\cite{PhysRevLett.82.2009,PhysRevA.58.4345,revzen:022103,quant-ph/0612029}.
Here we derive a BI that is based on Weyl symbols
for a large class of observables $\widehat{A}$ with dichotomic symbols.
As there is no general relation between
bounds of operators and bounds of their symbols, the resulting BI
is conceptually different than other BIs.

{\it Local Hidden-Variable Theories---}
In the context of dichotomic observables such as photon polarization, LHV theories
are based on the following assumptions.
\\
{\em LHV1}: the possible measurement values of a LHV observable $B$
are given by the spectrum of the corresponding quantum observable
$\widehat{B}$.
\\
{\em LHV2}: the expectation value of $B$ is given by the expression
$  \langle B \rangle_\text{LHV} =
  \int d\lambda \int db\, b\, p_\lambda(b)$,
where the integration (or summation) variable $b$ runs over the
spectrum of $\widehat{B}$. Here,
$\lambda$ represents all hidden variables in the LHV theory, and
$0\leq  p_\lambda(b) \leq 1$ denotes the probability to measure the
value $b$ if the hidden variable takes the value $\lambda$.
Conservation of probability requires that
$ \int d\lambda \int db\, p_\lambda(b)=1$.
\\
{\em LHV3}: For simultaneous measurements on two
spatially separated systems, causality implies that the probabilities
for the measurements on the two systems must be independent
(i.e., they usually factorize).

In the context of CV observables, the assumption LHV1 is problematic.
To see this, consider the case of a single 1D particle.
If we choose $\widehat{B}$ to be the position $\hat{q}$  or
momentum
$\hat{p}$ of the
particle, then the spectrum of $\widehat{B}$ is the set of real numbers.
Let us instead choose
$\widehat{B} = \widehat H = \frac 12 (\hat{q}^{2}+
\hat{p}^{2})$,  which
is the energy of a harmonic oscillator.
If, in a LHV theory, $q$ and $p$ can take any real value, then
the LHV spectrum of $B$ should include all positive real numbers.
However, the quantum mechanical energy spectrum is discrete.
Hence, if we tried to describe
position and energy simultaneously, we could run into a contradiction
about the spectra of the two observables.

The origin of this contradiction is of course that, in quantum
mechanics,  $\widehat{H}$ and $\hat{q}$ do not commute and thus cannot be
measured simultaneously. 
On the other hand, in LHV models the observables must be commutative
for at least a very large class of models
\cite{PhysRevA.69.022118}. The derivation of our generalized BI
is based on the assumption of commutativity.

{\it Re-derivation of the CHSH inequality.---}
Our method to derive BIs is related
to the proof of the CHSH inequality
presented by CFRD \cite{PhysRevLett.99.210405}, which
employs that
the variance of a (generally complex) observable
$B$ must be positive, $ | \langle B \rangle |^2 \leq \langle |B|^2
\rangle $. The expectation values in this expression may either
be evaluated within QM, then denoted by
$ \langle \widehat{B} \rangle $,
or in the framework of LHV models where we use the notation
$\langle B \rangle_\text{LHV}$. Within each theory this
inequality is always fulfilled. However, the maximum value of
$\langle |B|^2\rangle_\text{LHV}$ in all LHV theories provides
an upper bound on local realism: if the quantum mechanical
expectation value does not fulfill the inequality
\begin{equation}
  | \langle \widehat{B} \rangle |^2 \leq  \max_\text{LHV}  \langle |B|^2 \rangle_\text{LHV}\; ,
\label{eq:newBellConjecture1}\end{equation}
then the predictions of QM are inconsistent with the assumptions
behind LHV models.

To derive the CHSH inequality
we choose the observable
$
  B= X_1 X_2 + X_1 Y_2 + Y_1 X_2 - Y_1 Y_2
$,
with $X_i, Y_i $ four dichotomous observables (so that
$X_i^2=Y_i^2 =1$) for two particles $i=1,2$.
In quantum mechanics 
one finds
$
     \langle \widehat{B}^2 \rangle  =  4 +  \langle [Y_1,X_1]
     \, [X_2 ,Y_2] \rangle
$.
CFRD then argue that in any LHV model the commutators must vanish,
which leads to the
CHSH inequality $ | \langle \widehat{B} \rangle | \leq 2$.
QM violates this inequality for a suitable choice of states and
observables.
Because of the above-mentioned commutativity of many LHV models,
this seems to be a suitable approach to deriving BIs.

{\it Bell inequalities in phase space. ---}
In Ref.~\cite{PhysRevLett.99.210405}, CFRD used the method
of commuting observables to find BIs for a more general form
of the Bell observable $B$.
Here we employ this approach to derive BIs for a system of $n$
degrees of freedom (e.g., $n$ one-dimensional particles)
in phase space. We consider an operator
$\widehat{B}$ that is a function of
position and momentum operators $\hat{q}_i, \hat{p}_i\; (i=1,\cdots
n)$ and will establish an upper bound for $\langle
\widehat{B} \rangle $ in LHV models by ignoring the
commutators between position and momentum. 
We represent $\widehat{B}$ by a sum of Weyl-ordered products,
\begin{equation}
  \widehat{B} = \sum_{\mu_1, \cdots, \mu_{2n}}
           B_{\mu_1 \cdots \mu_{2n}}
           \left ( \hat{q}_1^{\mu_1} \hat{p}_1^{\mu_2}\cdots
           \hat{q}_n^{\mu_{2n-1}}\hat{p}_n^{\mu_{2n}} \right )_\text{Weyl}\; ,
\label{eq:Fexpansion}\end{equation}
with complex coefficients $B_{\mu_1 \cdots \nu_n} $.
Weyl ordering of a product of operators $\widehat{A}_1,
\widehat{A}_2 , \cdots$ corresponds to the completely symmetric sum
\begin{align}
 (\widehat{A}_1 \widehat{A}_2 \cdots\widehat{A}_k )_\text{Weyl}
 &:=
  \frac{ 1}{k!} \sum_P \widehat{A}_{P_1} \widehat{A}_{P_2} \cdots \widehat{A}_{P_k} \; ,
\end{align}
where $\sum_P$ stands for the sum over all permutations of the $k$ operators.
Any operator that possesses a Taylor expansion can be brought into the form
(\ref{eq:Fexpansion}). For instance, the operator $\widehat{B} =
\hat{q}\hat{p}$ can also be written as
$\widehat{B} = \frac{ 1}{2} (\hat{q}\hat{p}+\hat{p}\hat{q})+ i \frac{
  \hbar}{2} $, which corresponds to a single degree of freedom ($n=1$)
and $B_{1,1}=1$, $B_{0,0}= i \frac{\hbar}{2} $ in Eq.~(\ref{eq:Fexpansion}).

In phase space, the left-hand side (l.h.s.)
of Eq.~(\ref{eq:newBellConjecture1})
can be evaluated using relation (\ref{eq:weylExpec}).
The Weyl symbol of a Weyl-ordered operator $\widehat{B}$
can be evaluated by replacing the operators $\hat{q}_i,
\hat{p}_i$ by the respective phase space variables,
$\text{Smb}[\widehat{B}](q_i,p_i)= B(q_i,p_i)$. For the operator
(\ref{eq:Fexpansion}), the symbol
$B(q_i,p_i)$ then has the explicit form
\begin{equation}
 B(q_i,p_i) = \sum_{\mu_1, \cdots, \mu_{2n}}
           B_{\mu_1 \cdots \mu_{2n}}
          q_1^{\mu_1} p_1^{\mu_2}\cdots
           q_n^{\mu_{2n-1}} p_n^{\mu_{2n}}  .
\label{eq:Bexpansion}\end{equation}

To find the right-hand side (r.h.s.) of Eq.~(\ref{eq:newBellConjecture1})
we evaluate $ \langle \widehat{B}  \widehat{B}^\dagger  \rangle$ in QM and eliminate
the contributions of all commutators between position and momentum
operators. In quantum phase space one has
\begin{align}
  \langle \widehat{B}  \widehat{B}^\dagger\rangle &=
 \int\prod_{j=1}^n dq_j\, dp_j\, W(q_i,p_i)\,
  (B\star B^*)(q_i,p_i)\; ,
\label{eq:weylExpec2}\end{align}
where the star product between two operator symbols captures
the non-commutativity and non-locality of quantum observables 
and is given by \cite{FAB72,OM95,KO3}
\begin{eqnarray}\label{eq:groene}
 f\star g(x) &=& \frac1{(\pi\hbar)^{2n}} \int\int d^{2n}y\, d^{2n}z f(y)\,g(z) \\  \nonumber
   &\ &  \times\exp \frac {2i}{\hbar} \Big(y\cdot J z + z\cdot Jx + x\cdot Jy \Big)\,.
\end{eqnarray}
Here, $x=(q,p) \in \mathds{R}^{2n}$ and
$ J =\big[\begin{array}{cc}0&-I\\I&0\end{array}\big]$, where $I$
denotes the $n$-dimensional identity matrix.

We now employ the central assumption that the LHV upper bound
can be found by eliminating all commutators between position and
momentum operators. In the Weyl symbol representation this
can easily be accomplished by taking the limit $\hbar \rightarrow 0$
in the star product $B\star B^*$ of Eq.~(\ref{eq:weylExpec2}).
For smooth, $\hbar$-independent $f,g$ the star product has the
expansion
$    f\star g(x) = f(x)g(x) + \frac{i\hbar}{2}\{f,g\}(x) + \O(\hbar^2)$,
where the $\hbar$-linear term is a Poisson bracket \cite{FAB72,OM95,KO3}.
Consequently, $B\star B^*$ in Eq.~(\ref{eq:weylExpec2}) is replaced by
the algebraic product $|B|^2$ so that a tentative new BI is given by
\begin{align}
  |\langle \widehat{B} \rangle |^2
  &\leq  \int \prod_{j=1}^n dq_j\; dp_j\;  | B(q_i,p_i)|^2 \,
    W(q_i,p_i)\; .
\label{eq:BellTentative}\end{align}

The r.h.s.~of Eq.~(\ref{eq:BellTentative}) does not yet provide the
correct upper bound for LHV theories because it depends on the Wigner
function of QM.
Intuitively one could fix this by replacing the quantum state
by a suitable phase space distribution $W_\text{LHV}(q_i,p_i)$ which is
compatible with the assumptions of LHV models.
However, general LHV models do not necessarily possess such
a phase space distribution.

Instead, one has to consider suitable Bell operators for which the
 r.h.s.~of Eq.~(\ref{eq:BellTentative}) can be evaluated without
referring to a specific state. A simple but relevant example is the
case when $ | B(q_i,p_i)|^2 = |B_0|^2 $ is constant. This happens if
 $ B(q_i,p_i) $ is a phase factor or a sign function, for instance.
Because of the normalization of the Wigner function
the BI then becomes
\begin{align}
  |\langle B(\hat{q}_i, \hat{p}_i) \rangle |^2
  &\leq     | B_0|^2  \; .
\label{eq:BellCV}\end{align}
This is the main result of our paper. We remark that
the dependence on the state has been removed here because
the integral of the Wigner function over the entire phase space
equals to the probability to find any measurement result, which is
unity both in QM and LHV theories. Consequently, $ | B_0|^2 $
can be interpreted as the upper bound for commutative
LHV theories.

We emphasize that the BI (\ref{eq:BellCV}) is generally different
from CFRD or CHSH type Bell inequalities. First, the upper bound
is independent of the quantum state, while
for the CFRD inequality the state dependence needs to be addressed.
This can be a rather subtle issue:
some entangled quantum states do have a positive Wigner function
\cite{FoundPhys36-546}
and positivity of the Wigner function is not sufficient to ensure consistency
with LHV models \cite{PhysRevA.79.014104}.
In fact, Revzen {\it et al.} \cite{revzen:022103} have shown that a
dichotomic CV BI can be violated with a non-negative
Wigner function.
Second, the upper bound is determined by the symbol
of the operator $\widehat{B}$ rather than its spectrum. This can
be exploited to find new
examples of BI violation.

{\it Example of Bell inequality violation. ---}
We consider a system of two harmonic oscillators,
which are prepared in the Bell state
$ |\psi_\text{Bell} \rangle = \frac{ 1}{\sqrt{2}} (|0 \rangle  \otimes
  |1 \rangle -|1 \rangle  \otimes
  |0 \rangle ) $.
We set $\hbar=1$ and measure lengths in units of the
ground state width of the oscillators so that
 $\langle q|m \rangle = e^{-\frac{q^2}{2}} H_m ( q)/(\sqrt{2^m m! \sqrt{\pi}})$,
with  $H_m(q)$ the Hermite polynomials.
If we introduce a phase space variable
$\mathbf{x}:=(q,p)$ with $\mathbf{x}^2=q^2+p^2$, and
employ center-of-mass variables
$\delta\mathbf{x}:=(\mathbf{x}_1-\mathbf{x}_2)/\sqrt{2}$ and
$\mathbf{x}_\text{c} :=(\mathbf{x}_1+\mathbf{x}_2)/\sqrt{2}$,
the Wigner function
of $ |\psi_\text{Bell} \rangle$ is given by
\begin{align}
  W_\text{Bell}(\mathbf{x}_c, \delta\mathbf{x}) &= \pi^{-2} e^{- (\mathbf{x}_c^2 +\delta\mathbf{x}^2)}
  \left (2 \delta\mathbf{x}^2 - 1 \right )\; .
\label{eq:WignerBell}\end{align}
This corresponds to the product $W_{00}(\mathbf{x}_c) W_{11}(\delta
\mathbf{x})$ of the ground state in the center-of-mass coordinates and first excited state in
the relative coordinates.
To exploit the negativity of the Wigner function we consider
a Bell operator that has the dichotomic Weyl symbol
 $
  B(\mathbf{x}_1.\mathbf{x}_2) =
  \text{sgn} \left (  2\,\delta\mathbf{x}^2 - 1 \right )
$,
which implies that the bound $|B_0|^2$ in BI (\ref{eq:BellCV}) is unity.
We remark that the assumption that the Weyl symbol is dichotomic is
different from the assumption that the spectrum of an operator is
dichotomic. There is no general relationship between the bounds for
an operator and the bounds for its symbol.

The l.h.s. of BI (\ref{eq:BellCV}) can be evaluated using Eq.~(\ref{eq:weylExpec}),
so that the BI turns into
\begin{equation}
 |\langle \widehat{B} \rangle |^2 =
  \left |  \int d^2 \delta\mathbf{x}\, d^2 \mathbf{x}_c\; W(\delta\mathbf{x}, \mathbf{x}_c)\,
     \text{sgn} \left (  2\,\delta\mathbf{x}^2 - 1 \right )  \right |^2
  \leq 1 \; .
\end{equation}

For the Bell state (\ref{eq:WignerBell}) we find
$ |\langle \widehat{B} \rangle | = \frac{ 4}{\sqrt{e}} -1 \approx 1.426$,
so that the generalized BI is violated in quantum mechanics.
This violation is slightly larger than
Cirel'son's bound  $\sqrt{2}$ for the ratio between the maximal
quantum mechanical value of $ |\langle \widehat{B} \rangle |$ and  the
bound of the CHSH inequality for LHV theories \cite{LMP4-93}.

{\it Generalization. ---}
The previous example of BI violation exhibits two special
features: (i) in center-of-mass
variables, the Wigner function $W$ takes a product form. The negativity
of $W$ depends on just the relative phase space variable
$\mathbf{\delta x}$. (ii) The Bell operator's symbol
$B(\mathbf{\delta x})$ exactly cancels the negativity of $W$, so that
$\langle \widehat{B} \rangle =
   \int d^2 \mathbf{\delta x} \left | W(\delta\mathbf{x}) \right |$.
These features can be readily generalized.

Consider a Wigner function $W(\mathbf{x}) $, with
$\mathbf{x}\in \mathds{R}^{2n}$, that is negative
on a measurable subset ${\cal E}_- \subset \mathds{R}^{2n}$
and positive on its complement ${\cal E}_+$. We introduce
characteristic symbols $\chi_\pm(\mathbf{x})$ that are unity
for $\mathbf{x}\in {\cal E}_\pm$ and zero otherwise. The symbol
of the Bell operator is then given by
$B(\mathbf{x}) = \chi_+(\mathbf{x})-\chi_-(\mathbf{x})$.
The general relation between an operator 
on Hilbert space,  $\mathcal{H} = L^2(\mathds{R}^n,\mathds{C})$, and its
Weyl symbol is given by
\begin{equation}
  \widehat{A} = \int d^{2n} \mathbf{x} \,
  \widehat{\Delta}(\mathbf{x})\,
  \text{Smb}[\widehat{A}\,] (\mathbf{x} ) \,  ,
\label{eq:opFromSymbol} \end{equation}
with the quantizer \cite{KO3} defined via its Dirac kernel
\begin{equation}
    \langle q'|\widehat{\Delta}(\mathbf{x})|q''\rangle =
    (\pi\hbar)^{-n}
  \delta\left(q'+q'' -2q\right)\, e^{\frac{  i}{\hbar} p\cdot(q'-q'')}.
\end{equation}
The two operators corresponding to the characteristic symbols are
therefore given by
$  \widehat{\chi}_\pm = \int_{{\cal E}_\pm} d^{2n} \mathbf{x}\;
   \widehat{\Delta}(\mathbf{x})
$.
Together, they form a partition of unity,
$\widehat{\mathds{1}}= \widehat{\chi}_+ + \widehat{\chi}_-$,
 and commute. The non-local character of the star product (\ref{eq:groene}) causes
${\chi}_+ \star {\chi}_- \neq 0$  and as a result the Bell
operator fulfills $\widehat{B}^2 \neq \hat{\mathds{1}}$.
This implies that $\widehat{B}$ is not a dichotomic operator,
even though its symbol is dichotomic, in the sense that $B(\mathbf{x})^2 =1$.

As a concrete example, we consider the special case
that ${\cal E}_-$  consists of a disk
of (dimensionless) radius $R$  around the origin of a 2D phase space.
Since $\chi_-(\mathbf{x}) \in L^2(\mathds{R}^2,\mathds{C})$
the corresponding operator $\widehat{\chi}_-$ is Hilbert-Schmidt and so has a discrete spectrum.
Furthermore, because of its spherical symmetry
$\widehat{\chi}_-$ commutes with the Hamiltonian of the harmonic
oscillator, so that
the eigenstates of $\widehat{\chi}_-$ are the
harmonic energy eigenstates $|m \rangle $.

The eigenvalues $\lambda_m(R)$ of
$\widehat\chi_-$ are conveniently computed from the fact that they are
the expectation values with respect to the state $|m\rangle$
\cite{PhysRevLett.83.3758}
\begin{equation}\label{TAO1}
    \lambda_m(R) = 
\frac 1{2\pi\hbar} \int d^2 \mathbf{x}\, \chi_-(\mathbf{x})\,
\text{Smb}\left [ |m\rangle\langle m| \right ](\mathbf{x})\,.
\end{equation}
The integral above is the evaluation of the trace 
$\text{Tr}\, \chi_- |m\rangle \langle m|$ via the
corresponding Weyl symbol. Using 
$ \text{Smb}\left [|m\rangle\langle m| \right ] (\mathbf{x}) =
2(-1)^m\,L_m(2\mathbf{x}^2 ) \exp(2\mathbf{x}^2 )$
where $L_m$ is the Laguerre function gives
\begin{equation}\label{TAO2}
    \lambda_m(R) = \int_0^{R^2} (-1)^m\, L_m(r^2)\, e^{-r^2} \, d r^2\,.
\end{equation}
By replacing the Laguerre polynomials with their
generating function (Eq.~(22.9.15) of Ref.~\cite{abramowitz}),
we can derive a generating function for the eigenvalues
of $\widehat{\chi}_-$,
\begin{align}
   G(t) &=   (t-1)^{-1} \left ( e^{R^2\frac{t-1}{t+1}}-1 \right ) .
\end{align}
The $m$th eigenvalue $\lambda_m$ is given by
the $m$th coefficient of the Taylor expansion of
$G(t)$ around $t=0$.

\begin{figure}
\begin{center}
\includegraphics[width=6.5cm]{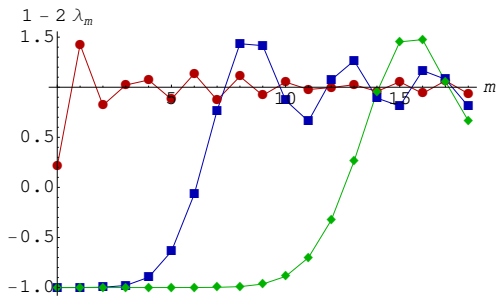}
\caption{\label{fig:BIviolation}
BI bound $|1-2\lambda_m| \leq 1$ for a system prepared in state
$|m \rangle $ and dichotomic operator symbols
corresponding to a disk of radius $R$ in phase space.
Shown are the cases $R=1/\sqrt{2}$
(circles), $R=3.5$ (squares),
and $R=5.5$ (diamonds).}
\end{center}
\end{figure}

For the Bell operator $\widehat{\mathds{1}}-2\widehat{\chi}_-$ and
the relative coordinate prepared in an eigenstate of
$\widehat{\chi}_-$, the BI then turns into the condition
$|1-2\lambda_m|^2 \leq 1$. In Fig.~\ref{fig:BIviolation} we
display the eigenvalues $1-2\lambda_m$ as a function
of the excitation number $m$ for different choices of $R$.
The example of the Bell state above corresponds to the case
$m=1$ and $R=1/\sqrt{2}$. For larger $R$ and higher excitation
numbers, a similar degree of BI violation can be obtained.

To increase the degree of BI violation, one can choose
${\cal E}_-$  to agree with the area in phase space where the
Wigner function $W_m$ of state $|m \rangle $ is negative, so
that $\langle \widehat{B} \rangle =
 \int d^2 \mathbf{x} \left | W_m(\mathbf{x}) \right |$.
\begin{figure}
\begin{center}
\includegraphics[width=6cm]{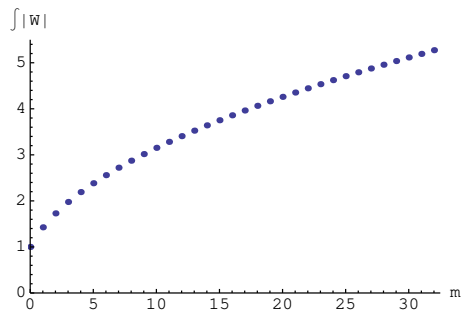}
\caption{\label{fig:Wnn}
BI violation of $|\widehat{B}| = \int d^2 \mathbf{x} \left | W_m(\mathbf{x}) \right |
\leq 1$, with $W_m$ the Wigner function of state $|m \rangle $.
 }
\end{center}
\end{figure}
The result for this expression for $m\leq 30$ is shown in
Fig.~\ref{fig:Wnn}. BI violations that are much larger than in the
CHSH case are possible. However, because of the normalization
and exponential decay of the Wigner function we conjecture
that the maximal BI violation for arbitrary $m$ will be bounded.

In conclusion, we have proposed generalized BIs for which the bound is determined
by the Weyl symbol of the Bell operator. Examples of states for which
the BI is strongly violated have been presented for bi-partite systems.
Extensions of this work for multi-partite or interacting systems may reveal
further insight into the differences
between classical and quantum mechanics.

\acknowledgments
This project was funded by NSERC and ACEnet.  
T.~A.~O. is grateful for an appointment as James Chair, and
K.-P.~M. for a UCR grant from St.~Francis Xavier University.

\bibliographystyle{apsrev4-1}
\bibliography{MoyalBell}
\end{document}